\documentclass{elsart}

\usepackage{natbib}

\usepackage{epsfig}

\usepackage{amssymb}

\begin{document}

\begin{frontmatter}



\title{Studies of Expolanets and Solar Systems with SPICA\thanksref{footnote1}}
\thanks[footnote1]{This template can be used for all publications in Advances in Space Research.}


\author{Michihiro Takami}
\address{Institute of Astronomy and Astrophysics, Academia Sinica, P.O. Box 23-141, Taipei 10617, Taiwan; phone +86-2-3365-2200 ext. 821; fax +886-2-2367-7849 }
\ead{hiro@asiaa.sinica.edu.tw}

\author{Motohide Tamura}
\address{National Astronomical Observatory of Japan, 2-21-1 Osawa, Mitaka, Tokyo 181-8588}
\ead{motohide.tamura@nao.ac.jp}

\author{Keigo Enya, Takafumi Ootsubo}
\address{Institute of Space and Astronautical Science, Japan Aerospace Exploration Agency, Yoshinodai 3-1-1, Sagamihara, Kanagawa 229-8510, Japan}
\ead{enya@ir.isas.jaxa.jp, ootsubo@ir.isas.jaxa.jp}

\author{Misato Fukagawa}
\address{Department of Earth and Space Science, Graduate School of Science, Osaka University, 1-1 Machikaneyama, Toyonaka, Osaka 560-0043, Japan}
\ead{misato@iral.ess.sci.osaka-u.ac.jp}

\author{Mitsuhiko Honda}
\address{Department of Information Science, Kanagawa University, 2946 Tsuchiya, Hiratsuka, Kanagawa 259-1293, Japan}
\ead{hondamt@kanagawa-u.ac.jp}

\author{Yoshiko Okamoto}
\address{Institute of Astrophysics and Planetary Sciences, Ibaraki University, 2-1-1 Bunkyo, Mito, Ibaraki 310-8512, Japan}
\ead{yokamoto@mx.ibaraki.ac.jp}

\author{Shigehisa Sako}
\address{Institute of Astronomy, University of Tokyo, Osawa 2-21-1, Mitaka, Tokyo 181-0015, Japan}
\ead{sako@ioa.s.u-tokyo.ac.jp}

\author{Takuya Yamashita}
\address{National Astronomical Observatory of Japan, 2-21-1 Osawa, Mitaka, Tokyo 181-8588}
\ead{takuya.yamashita@nao.ac.jp}

\author{Sunao Hasegawa, Hirokazu Kataza, Hideo Matsuhara,  Takao Nakagawa}
\address{Institute of Space and Astronautical Science, Japan Aerospace Exploration Agency, Yoshinodai 3-1-1, Sagamihara, Kanagawa 229-8510, Japan}
\ead{hasehase@isas.jaxa.jp, kataza@ir.isas.jaxa.jp, maruma@ir.isas.jaxa.jp, nakagawa@ir.isas.jaxa.jp}

\author{Javier R. Goicoechea}
\address{Centro de Astrobiologia (CSIC-INTA), 28850, Madrid, Spain}
\ead{goicoechea@damir.iem.csic.es}

\author{Kate Isaak}
\address{Cardiff University, The Parade, Cardiff CF243AA, UK}
\ead{Kate.Isaak@astro.cf.ac.uk}

\author{Bruce Swinyard}
\address{Rutherford Appleton Laboratory, Didcot, Oxfordshire, OX11 0QX, UK}
\ead{bruce.swinyard@stfc.ac.uk}

\begin{abstract}
The SPace Infrared telescope for Cosmology and Astrophysics (SPICA) is
a proposed mid-to-far infrared (4-200 $\mu$m) astronomy mission, scheduled for launch
in 2017.  A single, 3.5m aperture telescope would provide superior image quality 
at 5-200 $\mu$m, and its very cold ($\sim$5 K) instrumentation would provide superior sensitivity
in the 25-200 $\mu$m wavelength regimes. This would provide a breakthrough opportunity for studies of
exoplanets, protoplanetary and debris disk, and small solar system
bodies. This paper summarizes the potential scientific impacts for the
proposed instrumentation.

\end{abstract}


\end{frontmatter}

\parindent=0.5 cm

\section{Overview of SPICA}

The SPace Infrared telescope for Cosmology and Astrophysics (SPICA) is
a proposed mission for mid-to-far infrared (MIR/FIR) astronomy,
consisting of a single 3.5-m aperture space telescope with cooled
($\sim$5 K) instrumentation (see Nakagawa 2008). This mission would
provide a significant step forward in detection sensitivity in the 4
to 200 $\mu$m wavelength regime, which would revolutionize our
understanding of how galaxies, stars and planets form, and how 
interactions between complex astrophysical processes have ultimately
led to the formation of our own Solar system and the emergence of life
on Earth.

The role of SPICA would complement other forthcoming space infrared
missions; Herschel and JWST.  Herschel will provide improved
capabilities at wavelengths longer than 57 $\mu$m, while its
relatively warm temperature will produce a modest improvement over the
Spitzer Space Telescope.  JWST will provide the best performance at
wavelengths shorter than 25 $\mu$m, and will have the highest angular
resolution. SPICA would (1) achieve the best sensitivity at 25--200
$\mu$m, due to its cold mirror ($\sim$5 K); (2) achieve a clean
point-spread function at 5--200
$\mu$m due to its non-segmented mirror, thereby
providing the best performance for high-contrast coronagraphy; and (3) fill the gap in wavelength coverage between
Herschel and JWST.

SPICA would offer a unique opportunity for studying exoplanets, planet
formation, circumstellar disks, and small bodies in the solar system.
Table 1 summarizes the instruments proposed to date. Those selected
for the mission well be chosen based on technical constraints
(weight, volume, heat dissipation etc.) and scientific impact.  In \S
2, we describe potential scientific achievements in the above areas of
research.  In \S 3 we briefly describe the project schedule, including
the selection process for the instruments.

\section{Potential Scientific Achievements}
\subsection{Exoplanets (I): direct detection using coronagraphy}

In the last decade more than 300 gas-giant planets have been detected
through measurements of the radial velocity of their parent stars and
photometric measurements of transiting events.
These have lead us to the discovery of explanets around 6--7 \% of nearby
main-sequence stars (see Udry et al. 2007, Charbonneau et al. 2007, and references therein).
However, these detections are biased in favor of large exoplanets
with small orbital radii and parent stars with late spectral
types. Consequently, a number of extrasolar planetary systems may
have been missed by present observational techniques and capabilities.

Direct detection is a highly desirable method for the
study of exoplanets.  This technique has been hampered by the
extremely high brightness contrast between planets and their
parent stars.  Extensive coronagraphic observations of nearby stars
and possible planet-forming systems have been performed by the {\it
  Hubble Space Telescope} ($HST$), Subaru, Keck, VLT and Gemini.
Unlike the radial velocity method, this technique is sensitive to
planets with large orbital radii ($\gg$30 AU) towards stars.
Technical improvements in recent years have pushed
detection limits to on order of a Jupiter mass, leading to the
discovery of several candidate gas giants (Lafreni\`{e}re et al. 2008;
Kalas et al. 2008; Marois et al. 2008; Lagrange et al. 2008).

Space IR coronagraphy at wavelengths $>$3.5 $\mu$m could dramatically
improve such studies as (1) observations using space telescopes are
free from speckle and thermal noise caused by telluric 
atmosphere; (2) observations in the IR ($>$3.5 $\mu$m) minimize the
brightness contrast with the parent star (e.g., Burrows et
al. 2004). The high sensitivity of SPICA would also permit IR
spectroscopic observations free from atmospheric effects, thereby
allowing to study the physical properties and chemical abundances
of planetary atmospheres. Such a combination of IR space coronagraphy
and spectroscopic capability has only been proposed for SPICA.

SPICA would offer coronagraphy with an even higher contrast than JWST,
since its non-segmented mirror would provide a clean point-spread
function. The combination of SPICA and a dedicated coronagraph would
reach a detection limit with a contrast of $10^{-6}$ (see Fig 1).  Furthermore,
the coronagraph for SPICA is designed to cover a wide wavelength range
in order to study a variety of molecular bands, including CH$_4$ at
7.7 $\mu$m, H$_2$O at 6.3 $\mu$m, and NH$_3$ at 6.1/10.3/10.6 $\mu$m.
This spectroscopic capability, when combined with the coronagraph,
would allow detailed observations of these features and the
determination of atmospheric temperature and composition with high
accuracy. See Tamura (2000) and Enya et al. (2008) for details.

\subsection{Exoplanets (II): Observations of Transiting Planets}

The number of extrasolar planets detected through the transiting
planet method has been steadily increasing since the first discovery.
To date, dozens have been confirmed by this technique. It is strongly
expected that the number of such planets
detected will continue to increase rapidly through SPICA's
launch. COROT, launched in 2006 December, is expected to detect many
transiting giant planets and several transiting super-Earths (Borde et
al. 2003).  Kepler, successfully launched in March 2009, is expected
to detect numerous transiting giant planets and hundreds of transiting
super-Earths (Basri et al. 2005). The statistics for transiting
planets will also continue to improve through the use of ground-based
observations. 

While most transiting planets have been observed using the first eclipse, infrared photometry of the secondary eclipse (i.e., the passage of an extrasolar planet behind the parent star) allows to measure radiation from the extrasolar planet itself. A few extrasolar planets have been observed using this method, providing the temperature of their atmosphere (e.g., Deming et al. 2005; Charbonneau et al. 2005).

Spectroscopic observations of transiting planets are important for the
determination of the physical conditions of planetary atmospheres. 
While the first eclipse allows to observe absorption features 
from the planetary atmosphere as the stellar light pass through it,
infrared spectroscopy of the second eclipse provides spectra of the
planetary radiation itself.
Transiting spectroscopy of HD 209458b with $HST$
revealed deep absorption in H I, O I and C II, suggesting the presence
of an extended and escaping upper atmosphere beyond the Roche lobe
(Vidal-Madjar et al. 2003, 2004). More recently, molecules such as
H$_2$O, CH$_4$, CO, and CO$_2$ have been successfully detected using
the Spitzer Space Telescope and $HST$ in one such transiting planet, HD
189733b (Grillmair et al. 2007; Swain et al. 2008, 2009).

New missions such as Corot and Kepler, are expected to increase the
number of known transiting planets. Therefore, follow-up spectroscopy
with an IR telescope will be extremely valuable.  The sensitivities of
Spitzer, AKARI and ground-based 8-m telescopes are not sufficient for
such observations, thus a space IR telescope with a large diameter
is needed. SPICA, together with JWST, would have space-based mid-IR
spectroscopic capabilities, allowing to extend such observations to
the determination of the radius, density, and atmospheric compositions
of newly discovered transiting planets.

\subsection{Protoplanetary Disks}

Exoplanetary systems and the solar system are believed to have formed in
circumstellar disks, which are ubiquitous towards pre-main sequence
stars. Testing planet formation theories will require the observation
of disk in the process of active planet formation.  Space IR
spectroscopy is a powerful tool for observing
gas and dust associated with these disks.

The gas comprises most of the initial disk mass and may consequently play
an important role in the formation and evolution of planetary systems,
allowing gravitational instabilities to occur (e.g., Boss 2003), or providing gas drag
on rocky materials (e.g., Kominani \& Ida 2002).  So far, observational studies of gas disks have
been conducted mainly through radio interferometry and ground-based
optical-IR spectroscopy. The former technique allows to observe
regions on a few hundred AU scale (see Dutrey et al. 2007 for review),
while the latter allows to observe regions within a few AU of the
central star (see Najita et al. 2007 for review). Recent advances in
mid-to-far IR spectroscopy allow to explore the gas at
intermediate radii from a star (1--30 AU), the key zone for the
formation of planetary systems like our own.  Such studies include:
(1) Spitzer spectroscopy of atomic ([Ne II], [Fe II] etc.; e.g.,
Lahuis et al. 2007) and molecular lines (H$_2$O, OH, HCN, C$_2$H$_2$,
CO$_2$ etc.; e.g., Carr \& Najita 2008; Salyk et al. 2008); and (2)
ground-based observations of [Ne II] and H$_2$ at 17 $\mu$m (Herczeg
et al. 2007; Bitner et al. 2008).

The mid-to-far IR spectrographs on SPICA would extend such
observations to an unprecedented sensitivity and spectral
coverage. The mid-IR high resolution spectrograph would be sensitive
to the profiles of various emission lines, leading to the
determination of the column density distribution and physical/chemical
conditions as a function of radius (see Fig 2). To facilitate this, its spectral
coverage is designed to observe a variety of emission lines. The
mid-IR medium resolution spectrograph and far-IR spectro-imager would
detect emission lines over a wider spectral range (10--200 $\mu$m),
probing the total mass of the gas. Once we determine these physical
parameters as a function of age, we would be then be able to determine
the dissipation timescale of the gas disk. Furthermore, these lines
are responsible for gas cooling, and the observations would reveal a
clear picture of the energy balance in disks. All of the above
physical parameters are of vital importance for testing
various theories of planet formation.

Dust is the major building block for solid material in planets,
and is the major constituent of the terrestrial planets and the cores
of giant-gaseous planets. In particular, water ice associated with
dust grains is responsible for a significant amount of the total dust
mass, and could be important for sustaining life in extrasolar
planetary systems. While IR spectroscopic observations have previously
been used for extensive studies of silicate and carbon in dust grains
(see e.g., van Boekel et al., 2005; Kessler-Silacci et al. 2005; Honda
et al. 2006), our understanding of water-ice in protoplanetary disks
is far from complete. Water ice in protoplanetary disks is not readily
observable; we must either observe emission features at 44/62 $\mu$m
or absorption features in a spatially resolved disk.  Both are
extremely challenging, and there have been only a limited number of
successful observations (Malfait et al. 1999; Honda et al. 2009).

Although observations with Herschel and ground-based coronagraphs will
improve such studies, the high sensitivity of SPICA would provide much
more dramatic advances. The far-IR spectro-imager could be used to
observe ice features at 44/62 $\mu$m for a number of protoplanetary
disks.  These features hold the key to identifying the nature of the
ice (either amorphous or crystalline), and are therefore useful for
determining the thermal (and thereby chemical) history of the disks.
The MIR
grism spectroscopy would allow us observe a variety of solid
features (CO$_2$ etc.) in a manner similar to JWST.

\subsection{Debris Disks}
A number of debris disks have been discovered since the initial IRAS
identification of an infrared excess from the A-type main-sequence
star, Vega.  While most pre-main sequence stars with low-to-intermediate masses host
a dust disk, a recent census by Spitzer Space Telescope suggests that
10--15 \% of nearby main sequence stars host such disks, independent
of spectral type (see Meyer et al. 2007 for review). Further studies
with Spitzer have marginally resolved the structure of the thermal FIR
emission in the nearest and brightest debris disks (e.g., Stapelfeldt
et al. 2004; Su et al. 2005).  There is growing evidence that such
dusty disks are formed from debris produced mainly though the
collisions of planetesimals in the process of planetary system
formation.  Since the scales of most of the observed debris disks
correspond to the Kuiper belt in our solar system, studies of these
targets are important for understanding the origin and diversity of
extrasolar planetary systems. More recently, the relevance of such
studies has been highlighted by the discovery of a few exoplanet
candidates towards stars which host bright debris disks (Kalas et
al. 2008;Marois et al. 2008; Lagrange et al. 2008).

While observations of debris disks are often made at far-IR and mm
wavelengths, observations at shorter wavelengths (optical to MIR) have
significant advantages for the study of their structure.  In
particular, a combination of spectroscopy and polarimetry have
revealed a non-uniform distribution of dust properties, providing
clues for understanding their evolution (e.g., Okamoto et al. 2004;
Tamura et al. 2006). Morphological studies have been used to
investigate interactions with a possible exoplanet, providing an upper
mass limit (e.g., Kalas et al. 2008). However, statistical studies
have been severely hampered by the brightness of the central star:
such optical to MIR observations have been carried out for only a
handful of debris disks.

SPICA would offer several different approaches to the study of debris
disks. First, broad-band imaging with the far-IR spectro-imager could
lead to the discovery of an even larger number of debris disks
associated with nearby stars. Based on the estimated sensitivity, we
would be able to detect ``extrasolar Kuiper belts'' with a mass
comparable to that in our own solar system around nearby stars.
Secondly, FIR imaging and MIR coronagraphic imaging would allow the
study of the geometrical and physical structure of bright debris disks
in detail.  The morphological information provided by such an
instrument would be used to study the origin of the diversity of
debris disks, and determine the presence or absence of large gas-giant
planets near such disks.  The field of view of these instruments
(2'$\times$2' for SAFARI, FIR; 1'$\times$1' for coronagraph, MIR) is
sufficient to cover such disks in a single frame.  Thirdly, MIR
spectroscopy at 10--40 $\mu$m would enable us to observe the spatial
distribution of a variety of silicate emission lines, probing the
ongoing dust replenishment as a function of radius in a variety of
debris disks. Finally, spectro-imaging of the 44/62 $\mu$m ice
features towards the brightest debris disks could be used to infer the
presence or absence of the ``snow-line'', the possible boundary
between terrestrial and gas-giant planets.

\subsection{Solar System Small Bodies}

Since the first discovery of Trans-neptunian objects (TNO) in the
outer solar system (Jewitt \& Luu 1993), more than a thousand such
objects have been detected. Indeed, some have a size comparable to Pluto
(e.g., Bertoldi et al. 2006; Brown et al. 2006), leading us to define
a new category of solar system bodies at the IAU General Assembly in 2006.
Ongoing observations at optical
wavelengths are expected to find an even larger number, 
particularly at high ecliptic latitudes.  These objects presumably
belong to a physico-chemically unaltered population of the solar
system.

The diameter and albedo of these objects will allow to investigate
the initial conditions and dynamical evolution of the solar system. In
order to determine the size, albedo and thermal inertia of unresolved
small bodies, both the visible and thermal infrared brightness have to
be measured.  In particular, the measurement of the spectral energy
distribution (SED), which peaks at $\sim$100 $\mu$m, dramatically
decreases the uncertainties in the determination of the
size and albedo. These observations require imaging or photometric
facilities at far-IR (30--300 $\mu$m) wavelengths.  Spitzer had only a
few photometric bands in this range (24/70/160 $\mu$m), while Herschel
will not be able to observe wavelengths shorter than 60 $\mu$m.

SPICA would be ideally suited to the observation of the SEDs
of TNOs with unprecedentedly high sensitivity and accuracy (see Fig 3). Such
observations of a large number of TNOs would probe the conditions of
the `Initial Solar Nebula' in much greater detail than previously
accomplished.  Furthermore, low-resolution spectroscopy of the 44/62
$\mu$m water ice features would facilitate the study of the water ice
content and thermal history of the outer solar system.


\section{Summary and Project Status}

SPICA would provide significant advances in the study of exoplanets,
protoplanetary disks, debris disks and small bodies in the solar
system.  Its imaging capability would be able to detect debris disks
with an unprecedented sensitivity. These capabilities would allow
to measure the total flux of small bodies in the solar system,
useful for the study of their size distribution and albedos. SPICA's
spectral capability would probe the gas and dust associated with
circumstellar disks, and the atmospheres of the transiting
planets. Its coronagraphic capability would lead to the discovery of a
number of exoplanets, and the subsequent investigation of their
atmospheres and the geometry and distribution of ice in the spatially
resolved circumstellar disks. Possible achievements with the
individual instruments are summarized in Fig 4.

The conceptual design of SPICA is currently underway, and we will
begin the definition phase of the mission in 2009. This phase, which will
include the final decisions for the instruments, will be completed in 
2011. 
\\ \\ {\it Acknowledgement} --- We are grateful to Dr. Yuri Aikawa
and Akira Kouchi for useful discussions. We thank Dr. Jennifer Karr for reading our manuscript, and the editor and anonymous referee for useful comments.

\clearpage

\begin{table}
\caption{Instruments Proposed for SPICA Mission}
\begin{tabular}{lcll}
\hline
Instrument& Wavelength & Performance\\
          & ($\mu$m)\\
\hline
MIR Imager (MIRACLE) & 4--40 & 6'$\times$6' FOV, pixel scale = 0".35 \\
                        &    &+ capability of grism spectroscopy\\
FIR Spectro-Imager (SAFARI)$^1$ & 35--210 & 2'$\times$2' FOV, pixel scale = TBD \\
                        &      & with $R$ up to $\sim$2000 ($\Delta \lambda \ge$0.018 $\mu$m)\\
Far-IR Spectrograph (BLISS)   & 38--430$^*$ & $R$=700 ($\Delta \lambda$=0.05--0.6 $\mu$m)\\
MIR High-Resolution Spectrograph$^2$ & 4--18 & $R \sim 3 \times 10^4$ ($\Delta \lambda$=0.001--0.005 $\mu$m)\\
MIR Medium-Resolution Spectrograph & 10--40 & $R \sim 1 \times 10^3$ ($\Delta \lambda$=0.01--0.04 $\mu$m)\\
Coronagraphic Camera \& Spectrograph$^3$ & 3.5--27$^*$ & dedicated coronagraphy \\
                                  &       & 1'$\times$1' FOV, pixel scale = 0".059 \\
                                  &       & $R$=20--200 ($\Delta \lambda$=0.02--1.4 $\mu$m) for spectroscopy\\
\hline
\end{tabular}
\label{table1}
$^1$ Swinyard et al. 2009;
$^2$ Kobayashi et al. 2008;$^3$ Enya et al. 2008\\
$^*$ The present design of the entire SPICA mission covers 4--200 $\mu$m, although possible extension is optionally discussed for these instruments for greater scientific impacts.
\end{table}

\begin{figure}
\label{figure1}
\begin{center}
\includegraphics*[width=14cm]{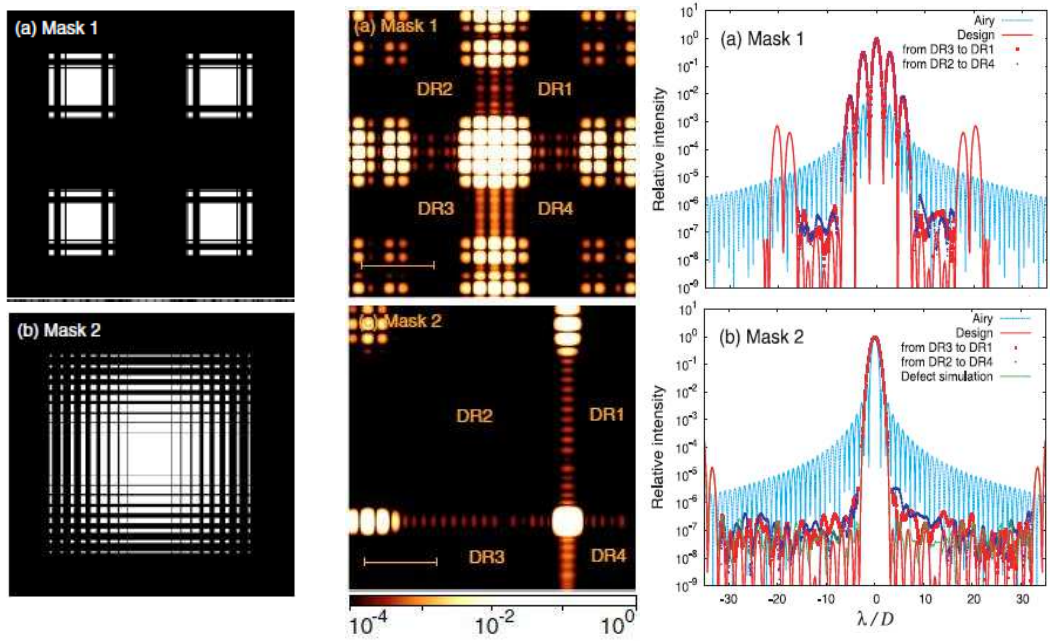}
\end{center}
\caption{Pupil mask images (left) and diffraction patterns obtained by experiments (middle and right). In the right figure, the Airy pattern of the telescope with a circular aperture is also shown. The special pupil mask dramatically improves the contrast of the point-spread function, in particular close to the stellar position. See Enya et al. (2007) for details.}
\end{figure}

\begin{figure}
\label{figure2}
\begin{center}
\includegraphics*[width=14cm]{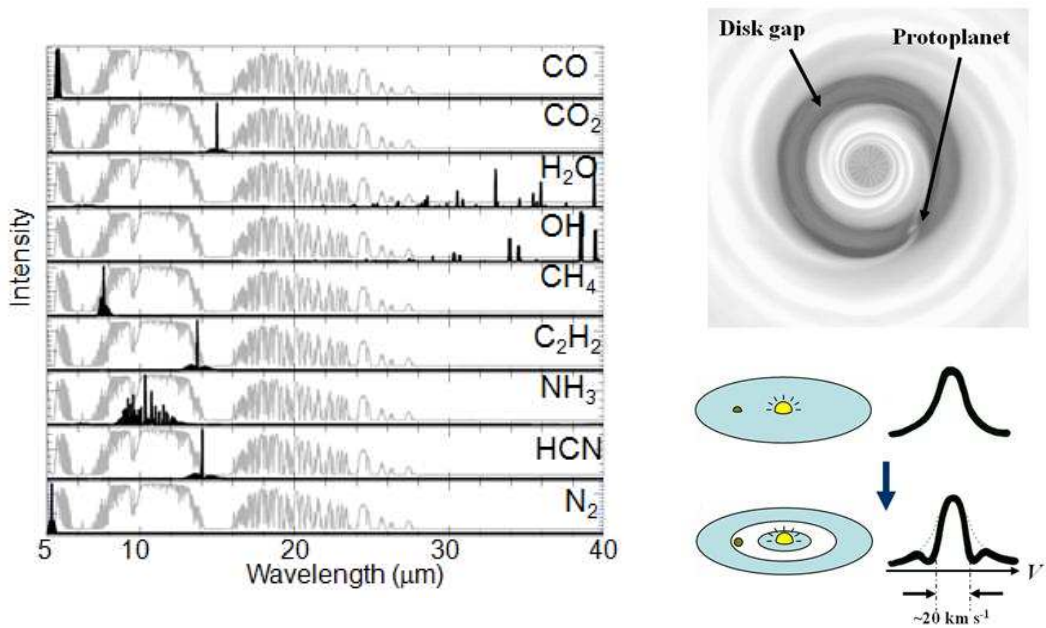}
\end{center}
\caption{(left) Band spectra of a variety of molecules at mid-IR wavelengths (optically-thin, 1000 K). (top-right) Numerical simulation for a protoplanet tidally interacting with a protoplanetary disk and opening-up a disk gap (Bryden et al. 1999). (bottom-right) Schematic view of how the disk clearing due to a proto-Jupiter would change the emission line profile.}
\end{figure}

\begin{figure}
\label{figure3}
\begin{center}
\includegraphics*[width=12cm]{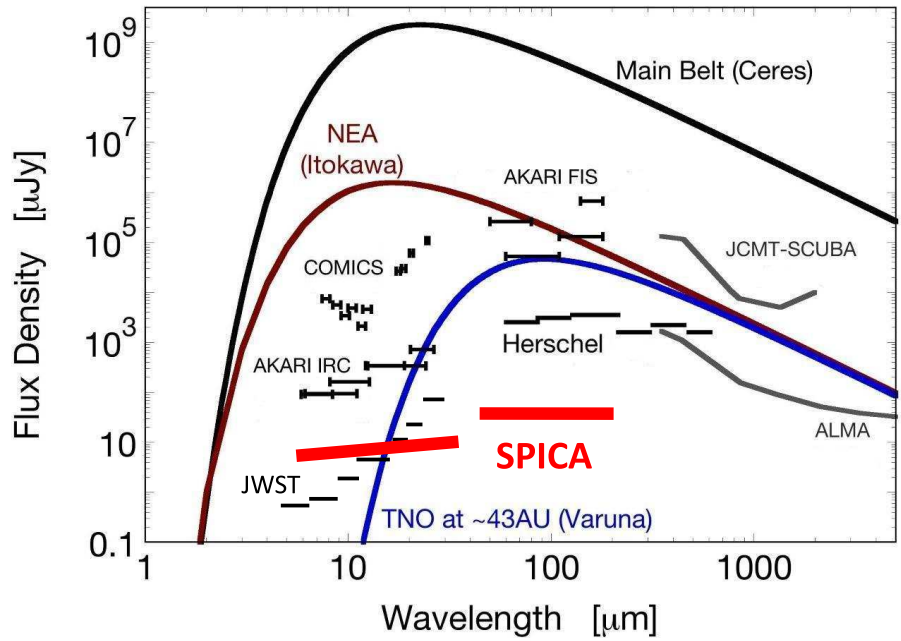}
\end{center}
\caption{Spectral energy distributions of asteroids (Ceres, Itokawa) and a Trans-neputunian object (Varuna), and detection limits of various facilities including Akari (5-$\sigma$, 1 pointing), Subaru-COMICS, JWST, Herschel, ALMA and SPICA (5-$\sigma$, 1 hour integration).
}
\end{figure}

\begin{figure}
\label{figure4}
\begin{center}
\includegraphics*[width=16cm]{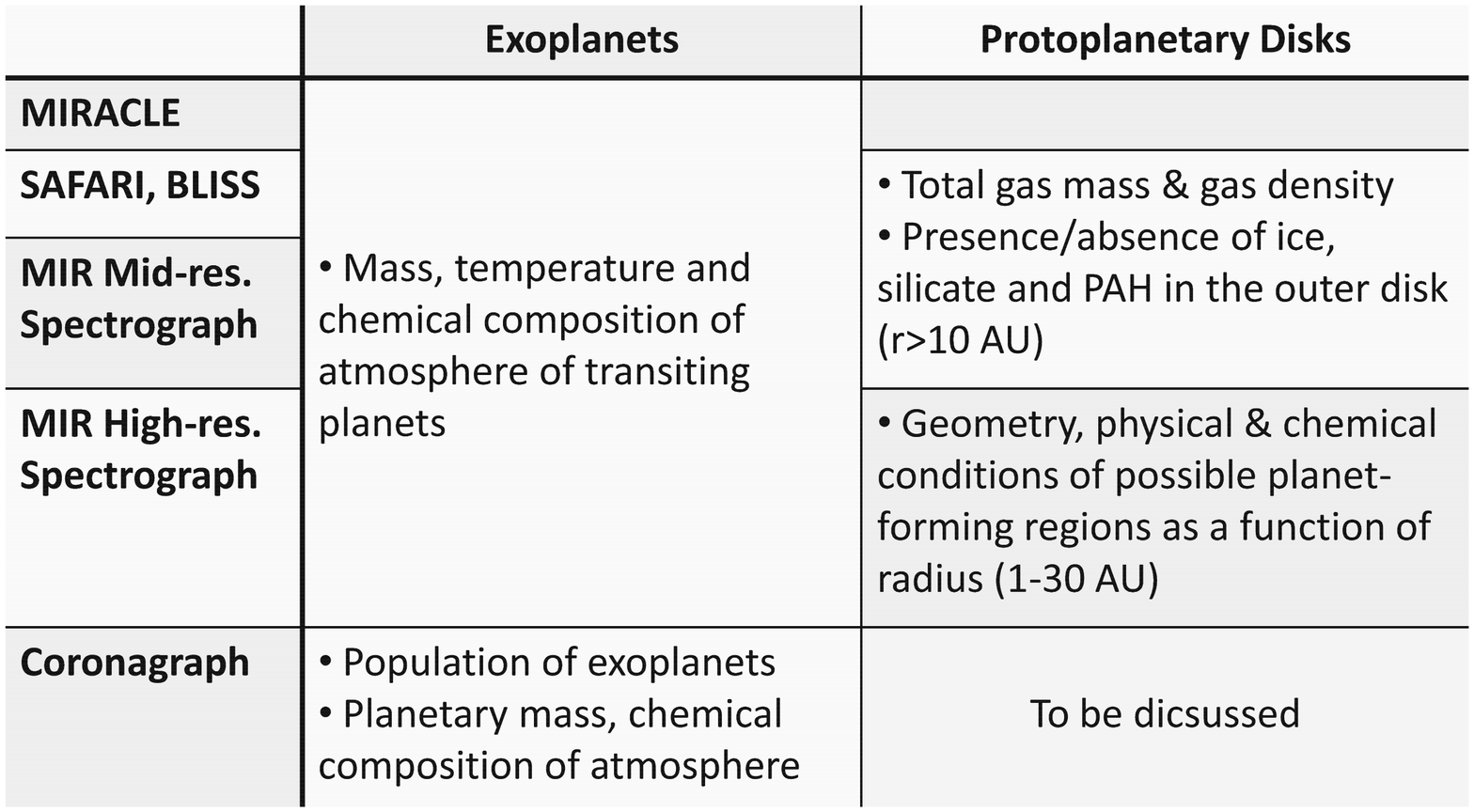}
\includegraphics*[width=16cm]{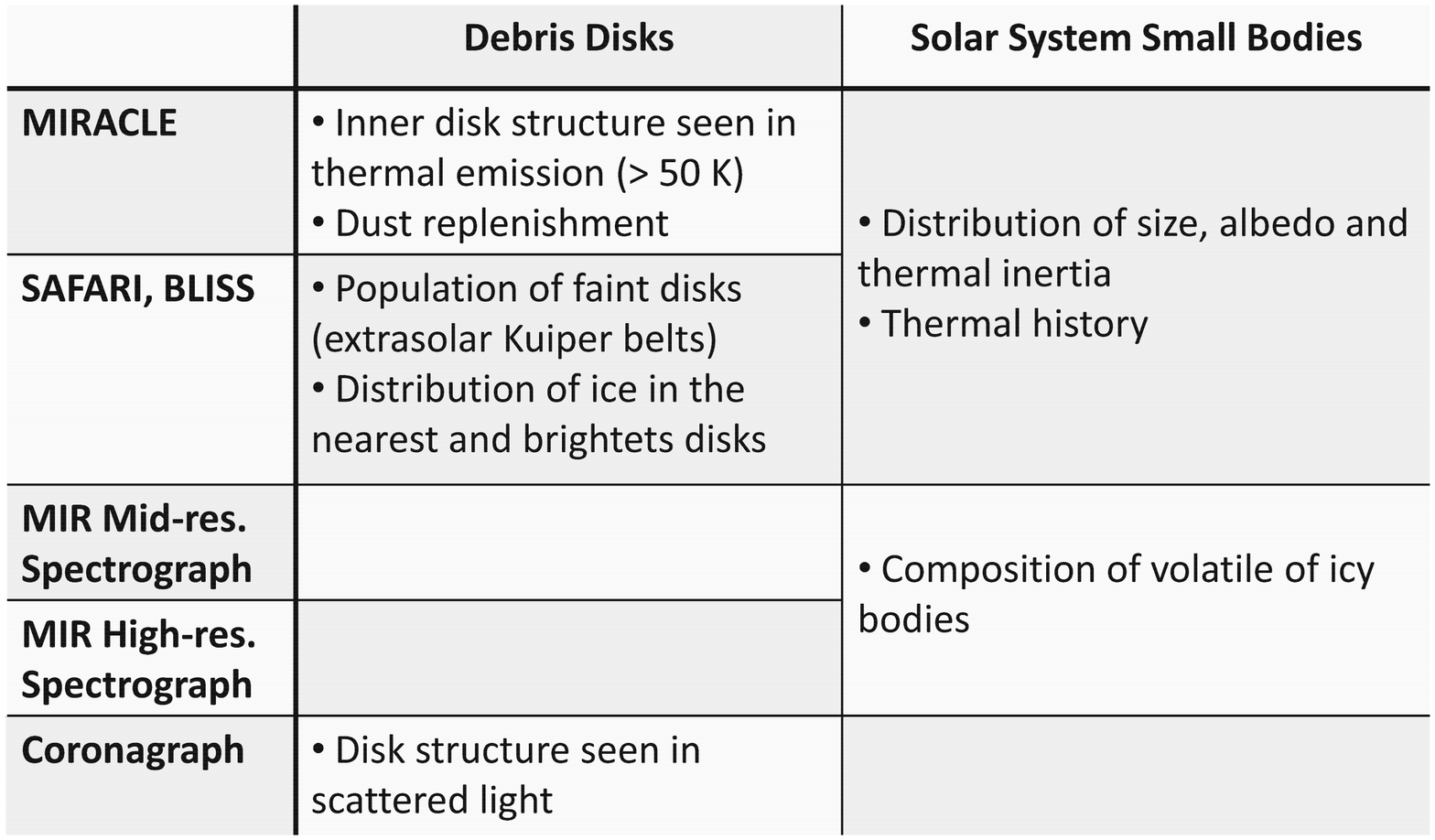}
\end{center}
\caption{Summary of parameters we will be able to measure on individual topics with different instruments}
\end{figure}

\end{document}